\documentclass[%
reprint,
 amsmath,amssymb,
 aps, prl
]{revtex4-1}

\usepackage{multirow}
\usepackage{graphicx}
\usepackage{dcolumn}
\usepackage{bm}
\usepackage{color}
\usepackage{fancyhdr}
\usepackage{textcomp}
\usepackage{soul}

\usepackage{hyperref}
\hypersetup{colorlinks=true, linkcolor=blue, citecolor=blue, urlcolor=magenta, pdftitle={Unveiling the room temperature magnetoelectricity of troilite FeS}, pdfauthor={F. Ricci, E. Bousquet}}

\begin{document}

\preprint{APS/123-QED}

\title{Unveiling the room temperature magnetoelectricity of troilite FeS}

\author{Fabio Ricci}
 \email{corresponding author: fabio.ricci@ulg.ac.be}
\author{Eric Bousquet}%
\affiliation{%
 Physique Th\'eorique des Mat\'eriaux, Universit\'e de Li\`ege, B-4000 Sart Tilman, Belgium
}%

\date{\today}

\begin{abstract}
The amazing possibility of magnetoelectric crystals to cross couple electric and magnetic properties without the need of time-dependent Maxwell's equations has attracted a lot of interest in material science.
This enthusiasm has re-emerged during the last decade where magnetoelectric and multiferroic crystals have captivated a tremendous number of studies, mostly driven by the quest of low-power-consumption spintronic devices.
While several new candidates have been discovered, the desirable magnetoelectric coupling at room temperature is still sparse and calls for new promising candidates.
Here we show from first-principles studies that the troilite phase of the iron sulfide based compounds, one of the most common mineral of Earth, Moon, Mars or meteors, is magnetoelectric up to temperatures as high as 415 K.

\end{abstract}

\pacs{Valid PACS appear here}

\maketitle



The troilite phase of FeS has been first depicted by an Italian Jesuite, Domenico Troili, during his analysis of a meteor that fells down in Italy in 1766 \cite{Troili1766}.
This particular phase among iron sulfide minerals is indeed commonly found in meteors originating from the Moon \cite{Muller1982} or Mars \cite{Martin2004} and it is also naturally found in earth crust \cite{Birch1952}, though most of them has a meteoritic origin.
Understanding the crystal properties of FeS is thus of high importance for planetary and geophysics studies such as planet's evolution
\cite{Allegre2001}. Numerous research investigations where focused on FeS in order to understand its complex temperature and pressure phase diagram \cite{Kusaba1998,Kusaba1997,ohfuji2007,ono2008}.
At high temperature, FeS is metallic and it crystallises in the the high symmetry hexagonal $P6_3/mmc$ space group, commonly called NiAs-type structure.
Below T$_N\sim$ 588 K, FeS undergoes an antiferromagnetic (AFM) phase transition with spins perpendicular to the hexagonal axis ($c$-axis) and around T$_s\sim$445 K a spin-flip transition occurs where the spins align toward the $c$-axis \cite{Adachi1968}.
At T$_\alpha =$ 415 K a structural phase transition arises changing the crystal structure from the $P6_3/mmc$ space group to the so-called troilite structure with space group $P\bar{6}2_c$ \cite{Andresen1960,Bertaut1980}.
Interestingly, the opening of an energy band gap accompanies this structural phase change such as FeS also experiences a metal-insulator transition at T$_{\alpha}$. Among the previous studies, there was a long debate, still unsolved today, whether the troilite phase of FeS is ferroelectric or not.
This argument was questioned because of the loss of the space inversion center symmetry during the troilite phase transition (the $P\bar{6}2_c$ space group is non-centrosymmetric) and also because electrical studies measured a ferroelectric polarization in this crystal, though no full evidence of ferroelectricity has been established \cite{Bertaut1956,Bertaut1997,Andresen1960,VandenBerg1969,VandenBerg1970,Li1996}.
We appealingly note that if FeS is ferroelectric then it would be a new candidate for room temperature multiferroism, a property that is nowadays highly desirable for multifunctional applications \cite{Scott2006,Khomskii2009}.
In addition, a very fresh study also observed a possible onset of a superconducting phase in troilite FeS from a meteoritic sample at a temperature as high as 117 K \cite{Guenon2015}.
All of that shows that FeS has unique multifunctional properties of tremendous potential for technological applications with the possibility to obtain samples with relatively cheap techniques directly from the Earth's crust and meteors.

In the present study, we cast light on the structural and multiferroic properties of trioilite FeS through first-principles studies.
The analysis of the phonon band structure of the high symmetry phase allows us to prove that FeS is not ferroelectric but it is nevertheless piezoelectric and magnetoelectric.
We show that the structural transition is driven by a zone boundary instability that couple to a zone center mode (similar to the trimerization observed in hexagonal manganites\cite{Cano2014}), which breaks the space inversion symmetry but without inducing an overall electric polarization.
While ferroelectricity is not induced, we prove that the crystallographic and AFM symmetries allow for magnetoelectricity.
Our analysis of the magnetoelectric properties shows that the amplitude of the response is about three times larger than Cr$_2$O$_3$ where the spin up and down channels can be seen as being electrically polarized in opposite direction. This later effect is at the source of magnetoelectric monopoles \cite{spaldin2013}.

\begin{figure*}[tb!]
\begin{center}
\includegraphics[scale=0.175,clip]{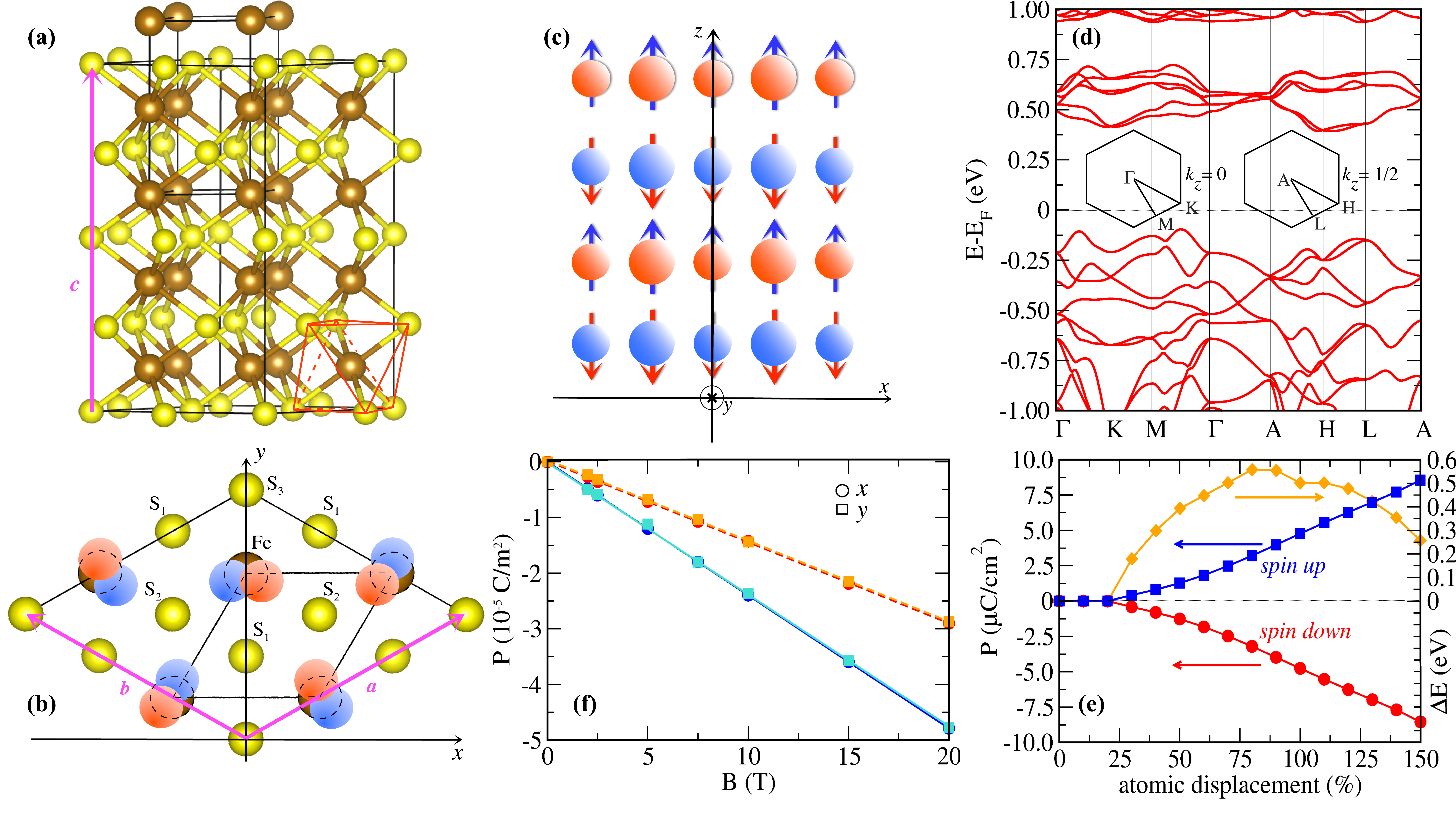}
\caption{(Color online) (a) Schematic view of the troilite cell (the brown spheres represent the Fe atoms and the yellow ones the S atoms) where the smaller cell defined by black lines is relative to the high symmetry $P6_3/mmc$ phase. The red lines highlight the octahedral coordination of one Fe.
(b) Top view of the troilite cell where the brown and yellow spheres are the Fe and S atoms in the high symmetry positions. 
The blue and red circles represent the displacement of the Fe atoms (the two colours distinguish the two different Fe planes along the $c$-axis) away from their initial positions (dashed circles). 
(c) Fe AFM arrangement: red and blue colours highlight different Fe spin planes (the S atoms are not shown).
(d) Electronic band structure for the $P\bar{6}2_c$ phase and the insets give a top view of the
$k_z=0$ and $k_z=1/2$ Brillouin Zone planes (see Fig. \ref{fig:002} for the point coordinates).
(e) Spin up (blue squares) and spin down (red circles) contributions to the electric polarisation and energy gap (orange diamonds) as a function of the fraction of the atomic distortions between the ground state $P\bar{6}2_c$ (100\%, highlighted by a vertical dashed line) and $P6_3/mmc$ (0\%) phases.
(f) Electric polarisation as a function of applied magnetic field along the $x$ (circles) and $y$ (squares) directions. Plain lines represent the total contribution and the dashed ones the electronic contribution.}
\label{fig:001}
\end{center}
\end{figure*}

--- \textbf{Results} ---
\textbf{Is FeS ferroelectric/multiferroic ?} 
In this first section we propose to elucidate the ferroelectric character of the room temperature phase of FeS.
The troilite phase of FeS is found experimentally to crystalise in the hexagonal $P\bar{6}2_c$ space group with 12 Fe atoms and 12 S atoms in the unit cell (see Fig. \ref{fig:001}(a) and (b)) and it is insulator with a small band gap of about 0.04 eV at room temperature \cite{Gosselin1976}.
The Fe atoms order antiferromagnetically with spins collinearly aligned along the $c$-axis where the magnetic structure can be seen as alternating planes of spin up and spin down along the $c$ direction (see Fig. \ref{fig:001}(c)).
Performing a full-cell relaxation from first-principles calculations without any Hubbard $U$ correction (see Methods Section) on the Fe-$d$ orbitals we recover the insulating state with the GGA exchange-correlation functionals but not with LDA.
Over the different approximations, we found that the GGA PBE functional with $U=1$ eV gives the best agreement with the experimental results (see Supplemental Material) for the structural parameters and we will present results within this approximation in the next unless stated otherwise.
We found that the ferromagnetic (FM) order is 10 meV higher in energy than the AFM one and our non-collinear calculations stabilises the out-of-plane alignment of the spins with no canting, which is in agreement with experimental observations \cite{Horwood1976} and with previous calculations \cite{Antonov2009}. 

In Fig. \ref{fig:001}(d), we report the electronic band structure of our relaxed $P\bar{6}2_c$ phase of FeS around the last occupied bands.
We found an indirect band gap of 0.49 eV, which is one order of magnitude larger than the experimental gap but in agreement with previous calculations
\cite{Rohrbach2003,Hobbs1999}.
We note that the overestimation of the band gap can originate from the exchange-correlation approximation used in our DFT simulations but also from experimental band gap underestimation due to the presence of defects or off stoichiometry \cite{VandenBerg1969,Gosselin1976}.
Regarding the character of the bands in Fig. \ref{fig:001}(c), we remark that around the band gap the bands are mostly of Fe-$d$ character and the S-$p$ bands have low weight (see Supplemental material).
This orbital character arrangement in the band structure, also found in the metallic FeS high symmetry phase (see below), is typical of iron-based superconductors \cite{Chen2014} and it might be at the source of the recent observation of a possible superconducting transition in troilite FeS \cite{Guenon2015}.
While we don't observe metallicity in the troilite FeS ground state, we argue that the source of a superconducting phase might come from vacancies and polarons as observed for example in WO$_3$ \cite{Salje}.

At this point, in order to understand whether this alloy has a ferroelectric porization in its ground state, we performed Berry phase calculations to compute the polarization in the $P\bar{6}2_c$ phase and, after taking care of the quantum of polarization, we found that the total polarization is zero.
However, when decomposing the spin channel electric polarization contributions we found an absolute value of about 5 $\mu$C/cm$^{2}$ for each spin direction with opposite sign such as the total polarization is zero.
To exemplify this effect, we computed (within the collinear scheme) the spin up and down electric polarization at different amplitudes of the pattern of displacements that drives FeS from the high symmetry $P6_3/mmc$ phase to the $P\bar{6}2_c$ phase and we plot the results in Fig. \ref{fig:001}(e).
Interestingly, from zero to about 25\% of the $P\bar{6}2_c$ distortion, no electric polarization develops in the two spin channels; The system being metallic (see Fig. \ref{fig:001}(e) for the gap evolution with amplitude of the distortion), it forbids any electric polarization onset.
Beyond 25\% of the total distortion the band gap opens and an electric polarization develops in each spin channel but in opposite directions.

Using the classical picture, the electric polarization corresponds to the integration of the charge times the position operator $\mathbf{r}$.
Within the DFT scheme, it can be expressed through the density times the position operator: $\rho(\mathbf{r})\cdot\mathbf{r}$, where $\rho(\mathbf{r})=\rho^\uparrow(\mathbf{r})+\rho^\downarrow(\mathbf{r})$, in terms of spin up $\rho^\uparrow(\mathbf{r})$ and spin down $\rho^\downarrow(\mathbf{r})$ contributions.
On the other hand, replacing the charge density $\rho(\mathbf{r})$ by the magnetization density $m(\mathbf{r})$=$(\rho^\uparrow(\mathbf{r})-\rho^\downarrow(\mathbf{r}))$, we obtain the definition of the magnetoelectric monopole $A$, which is given by the space integration of $m(\mathbf{r})\cdot\mathbf{r}$ as for the electric polarization \cite{spaldin2013}.
Interestingly, we see that in FeS the electric polarization is zero but it has a non-zero magnetoelectric monopolarization $A=$5.9$\ 10^{-3}$ $\mu_B$/\AA$^2$, which means that it is not ferroelectric but it is nevertheless magnetoelectric.

Magnetoelectricty is a spin-orbit driven crystal response (when disregarding the exchange-striction effects) and it is observed either by a magnetic field induced electric polarization or by an electric field induced magnetization \cite{fiebig2005}.
In order to estimate the amplitude of the magnetoelectric response, we made non-collinear calculations with spin-orbit interaction where we performed a full relaxation under a finite magnetic field and then we extracted the induced electric polarization (see Method section).
We present these results in Fig. \ref{fig:001}(f) where we plot the total induced polarization and its electronic contribution versus the amplitude of the magnetic field.
The electronic contribution is obtained by applying the field without letting the atoms to relax; In this way, only the electrons respond to the field (clamped ions or high frequency response \cite{bousquet2011}).
From Fig. \ref{fig:001}(f), we see that applied magnetic fields along the $x$ and $y$ directions induce an electric polarization in the same direction while we do not see any induced polarization when the field is applied in the $z$ direction (not shown). 
This can be understood from simple arguments: The system being collinear with spins along the $z$ direction, at 0 K the application of a Zeeman magnetic field parallel to the magnetic moments will not induce any response beside a phase transition form AFM to FM order at large field amplitudes.
In addition, the induced responses along the $x$ and $y$ directions have the same amplitude (we show the response with a field along the $x$ direction only on Fig. \ref{fig:001}(f) for clarity) and are linear.
The magnetoelectric response is thus simply given by the slope of these curves and we found $\alpha_{xx}^{tot}\simeq\alpha_{yy}^{tot} =$ 3.00 ps/m. We checked the $U$ and $J$ parameters dependence of $\alpha_{xx}$ and $\alpha_{yy}$ and we observed values going from 3 to 6 ps/m (see Supplemental Material). 
The magnetoelectric response of FeS is thus about two to three times larger than the one reported for Cr$_2$O$_3$ \cite{iniguez2008, bousquet2011, malashevich2012}.
The electronic contribution to the response gives $\alpha_{xx}^{el}\simeq\alpha_{yy}^{el} =$ 1.83 ps/m, which represents 61\% of the total response.
The electronic magnetoelectric response is thus large in FeS and it is more important than the one reported in Cr$_2$O$_3$ where the electronic contribution represents about 25\% of the total response \cite{bousquet2011}.

Other related quantities to magnetoelectric crystals are the magnetic effective charges, which represent the change of magnetisation against atomic displacement \cite{iniguez2008,Vanderbilt2014}. We calculated these magnetic effective charges in FeS and we found that the largest components are Z$^M_{xx}$=Z$^M_{xz}$=-Z$^M_{yx}$=$0.1\ \mu_B$/\AA. Interestingly, these values are two times larger than those reported in Cr$_2$O$_3$ \cite{Vanderbilt2014} and thus in agreement with the amplitude of the magnetoelectric response. 


We note that a previous study of Li \emph{et al.} \cite{Li1996} proposed to attribute the ferroelectric phase of FeS as being a polar subgroup of the $P\bar{6}2_c$ phase ($P31c$ polar group) in which a polar mode would induce a ferroelectric phase transition in the troilite phase.
We explored the possibility of a $P31c$ polar phase by performing two tests: (i) Condensing the polar mode in the $P\bar{6}2_c$ and relaxing the structure to see whether a lower energy polar phase can be reached and (ii) Performing phonon calculations in the $P\bar{6}2_c$ phase to see if a polar unstable mode is present. We find that both tests contraindicate the existance of the $P31c$ polar phase. 
In (i) the full relaxation drives the system back to the $P\bar{6}2_c$ phase without any remaining polarization or gain of energy.
In (ii) we do not find any unstable nor soft mode in the $P\bar{6}2_c$ phase.
These tests show that the troilite phase is at least locally stable against atomic and strain distortions.

From our DFT calculations we were thus able to conclude that the troilite phase of FeS is not ferroelectric but it is nevertheless a new room temperature magnetoelectric candidate, a sought after property of the last decade \cite{fiebig2005}.

--- \textbf{Microscopic origin and symmetry analysis} ---
To understand the phase transitions that occur in FeS and the microscopic origin of its magnetoelectric phase, we analysed the electronic and vibrational properties of the high temperature $P6_3/mmc$ phase.
As for the low symmetry phase, the GGA PBE approximation with $U=$ 1 eV gives the best agreement on the relaxed structure against the experimental values (see Supplemental Material).

\begin{figure*}[tb!]
\begin{center}
\includegraphics[scale=0.26,clip]{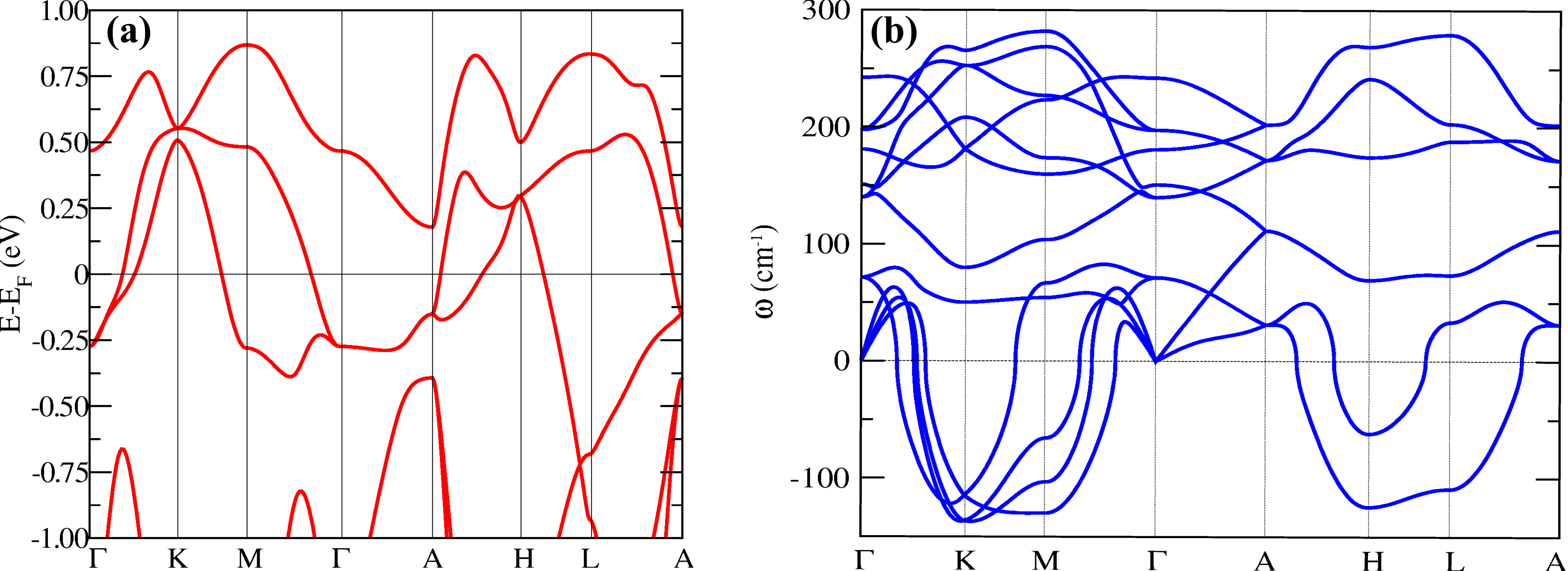}
\caption{(Color online) (a) Electronic band structure and (b) phonon dispersion curves including total and atom-projected density of states of the high symmetry $P6_3/mmc$ FeS phase. The special k-points in the hexagonal Brillouin Zone are (in reduced coordinates): $\Gamma\ (0,0,0)$, K $\left(\frac{1}{3},\frac{1}{3},0\right)$, M $\left(\frac{1}{2},0,0\right)$, A $\left(0,0,\frac{1}{2}\right)$, H $\left(\frac{1}{3},\frac{1}{3},\frac{1}{2}\right)$, L $\left(\frac{1}{2},0,\frac{1}{2}\right)$.}
\label{fig:002}
\end{center}
\end{figure*}

In Fig. \ref{fig:002}(a) we report the electronic band structure of the $P6_3/mmc$ phase where we recover the aforementioned property of mostly Fe-$d$ character of the bands around the Fermi level, with the interesting difference that the system is metallic as observed experimentally at high temperature in this phase.
In Fig. \ref{fig:002}(b) we report the phonon dispersion curves of the metallic $P6_3/mmc$ phase.
These dispersions show that the high temperature phase presents several unstable phonon branches (imaginary frequencies plotted as negative values on Fig. \ref{fig:002}(b)) at the zone boundary K, M, H and L points.
This means that the $P6_3/mmc$ phase is unstable over several type of atomic patterns of distortions.
The strongest instabilities are observed at the K point with an irreducible representation (irrep.) K$_5$ (138$i$ cm$^{-1}$), at the M point with M$^-_{2}$ irrep. (131$i$ cm$^{-1}$), at the H point with H$_2$ irrep. (126$i$ cm$^{-1}$) and at the L point with L$_1$ irrep. (111$i$ cm$^{-1}$).
We remark that the H$_2$ irrep. gives the sub-group $P\bar{6}2_c$ (see Tab. \ref{tab:001}), which is the ground state space group of the low temperature magnetoelectric phase.
This suggests that the condensation of the H$_2$ unstable mode in the $P6_3/mmc$ phase would drive directly the system to the ground state phase.
We thus condensed the H$_2$ mode in the $P6_3/mmc$ phase, performed a full relaxation of the cell and we indeed obtained the $P\bar{6}2_c$ ground state phase with the same energy as the one we discussed in the previous section (confirming that the two phases are identical).
We would thus a priori conclude that the H$_2$ mode alone is at the origin of the troilite phase.
To check this hypothesis, we performed a mode decomposition of the $P\bar{6}2_c$ relaxed phase with respect to the phonon mode basis of the high symmetry $P6_3/mmc$ phase using the AMPLIMODES software \cite{amplimode} and we present the results in Tab. \ref{tab:001}.
Interestingly, we observe that the H$_2$ mode is certainly contributing the most to the distortions (1.46 \AA) but we also have the unstable K$_5$ mode and the stable $\Gamma^-_{4}$ mode that contribute for 0.53 \AA\ and 0.18  \AA, respectively.
This indicates that the $P\bar{6}2_c$ is driven by the H$_2$ unstable mode but its condensation allows the development of the additional K$_5$ and $\Gamma^-_{4}$ modes.
These mode combinations can be explained by symmetry analysis: If one expands the energy with respect to the H$_2$, K$_5$ and $\Gamma^-_{4}$ mode distortions, he has to respect the invariance of the energy with respect to the symmetries of the $P6_3/mmc$ phase. If we consider the H$_2$ domain reported in Tab. \ref{tab:001}, we found the terms $a\times b^2$ and $c\times b^2$ at the third order with $a$ and $c$ the amplitudes of the $\Gamma_4^-$ and K$_5$, respectively (see Tab. \ref{tab:001}). Then, the H$_2$ mode drives through improper-like coupling the $\Gamma^-_{4}$ and K$_5$ modes, in a similar way to the trimerization observed in YMnO$_3$ \cite{Fennie2005,Cano2014} or in MoS$_2$ \cite{Shirodkar2014}.

We note that $\Gamma^-_{4}$ is a mode that breaks the space inversion symmetry but it is not infra-red active and it is thus non-polar, confirming that the $P\bar{6}2_c$ phase is not ferroelectric.
Regarding the magnetic space group, we obtain the $P\bar{6}'2c'$ magnetic point group when the spins lie along the $c$ direction, which allows for a diagonal magnetoelectric tensor with $\alpha_{xx}=\alpha_{yy}\neq\alpha_{zz}$.
Then, the symmetry analysis confirms that the troilite phase is magnetoelectric as obtained in our DFT calculations.

\begin{table*}
\begin{center}
\begin{tabular}{|c|c|c|c|c|}
\hline
    $k$-vector & Irrep. & Direction & Subgroup &  Amplitude (\AA)\rule{0pt}{0.30cm} \\
\hline
 (0,0,0) & $\Gamma^-_{4}$ & $(a)$ & $P\bar{6}m2$ (187) & 0.18 \rule{0pt}{0.40cm} \\[1ex]
($\frac{1}{3}$,$\frac{1}{3}$,$\frac{1}{2}$) & H$_2$ & $(-\sqrt{2}/{2}\ b,-1/2\ b,0,0)$ & $P\bar{6}2_c$ (190)& 1.46 \rule{0pt}{0.40cm}\\[1ex]
($\frac{1}{3}$,$\frac{1}{3}$,0) &  K$_5$ & $(0,0,-1/2\ c,-\sqrt{2}/{2}\ c)$ & $P\bar{6}2_m$ (189) & 0.53 \rule{0pt}{0.40cm}\\[1ex]
\hline
\end{tabular}
\caption{Symmetry adapted modes decomposition of the relaxed $P\bar{6}2_c$ phase with respect to the $P6_3/mmc$ phase as obtained from the {\it AMPLIMODES}\ software \cite{amplimode}. In the first column we show the $k$-vector coordinates, in the second one the Irrep. of the symmetry adapted mode, in the third one the direction of the mode condensation, in the fourth one the corresponding subgroup, in the fifth one the amplitude of the mode distorsion.
}
\label{tab:001}
\end{center} 
\end{table*}

--- \textbf{Discussion} ---
In spite of the last decade effort in the search of magnetoelectric materials influenced by the exciting possibilities brought by spintronic applications, we are still facing a scarcity of room temperature crystal candidates \cite{Fiebig2009}.
While most studies in the field of magnetoelectrics focused on oxide materials, here we showed that the common iron sulphide troilite mineral found on Earth, Moon, Mars and meteors is magnetoelectric up to $\sim$415 K.
Our DFT calculations give that the amplitude of the magnetoelectric response of FeS is of the same order of magnitude than Cr$_2$O$_3$.
We also showed that the room temperature phase comes from a displacive phase transition in which a zone boundary soft mode condenses in the high temperature metallic phase, driving the opening of the band gap.
Additionally, our electronic structure analysis pointed toward a similar electronic structure character of FeS as the one observed in the Fe-based superconductors, that might explain the recent observation of a possible superconducting phase at 117 K. 
These results are thus of primary importance in a widespread field of research, going from Earth and planetary studies to multifunctional applications and further theoretical and experimental studies of FeS are highly appealing.

--- \textbf{Methods} ---
We performed density functional theory (DFT) {\em ab-initio} simulations using the ABINIT \cite{Gonze2009} and VASP \cite{vasp1,kresse1999} codes within the local density approximation (LDA) and the PBE \cite{Perdew1996} and PBEsol \cite{Perdew2008} flavours of the generalized gradient approximation (GGA).
We also make use of the DFT$+U$ correction in order to treat the localised $d$ orbitals of Fe atom \cite{Liechtenstein1995}.
We used projected augmented-wave (PAW) pseudopotentials \cite{Torrent2008} and in order to achieve a satisfactory degree of convergence ($\sim$0.01 meV energy differences) the plane wave expansion has been truncated at a cutoff energy of 550 eV and the integrations over the Brillouin Zone was performed considering 16$\times$16$\times$10 and 10$\times$10$\times$6 uniform Monkhorst and Pack \cite{Monkhorst1976} grids for the high symmetry $P6_3/mmc$ and low symmetry $P\bar{6}2_c$ unit cells, respectively. 
We calculated the phonon dispersions using the density functional perturbation theory \cite{gonze1995b,gonze1997}, the electric polarization through the so-called Berry phase technique \cite{kingsmith1993}. Non-collinear magnetism has been performed by including the spin-orbit coupling.
The magnetoelectric response has been calculated with an applied Zeeman magnetic field \cite{bousquet2011} and using the LDA$+U$ functional at the the GGA$+U$ relaxed cell as described in Ref. \cite{bousquet2011b}.

--- \textbf{Aknowledgements} --- 
FR and EB thanks the FRS-FNRS, the Consortium des Equipements de Calcul Intensif (CECI) funded by the FRS-FNRS (grant 2.5020.11) and the PRACE project TheDeNoMo. We also thank N. Spaldin and F. Th\"ole for useful discussion regarding the magnetic monopolarization and A. Cano, Ph. Ghosez and M. Verstraete for their fruitful help in the symmetry analysis.

%


\begin{thebibliography}{50}%
\makeatletter
\providecommand \@ifxundefined [1]{%
 \@ifx{#1\undefined}
}%
\providecommand \@ifnum [1]{%
 \ifnum #1\expandafter \@firstoftwo
 \else \expandafter \@secondoftwo
 \fi
}%
\providecommand \@ifx [1]{%
 \ifx #1\expandafter \@firstoftwo
 \else \expandafter \@secondoftwo
 \fi
}%
\providecommand \natexlab [1]{#1}%
\providecommand \enquote  [1]{``#1''}%
\providecommand \bibnamefont  [1]{#1}%
\providecommand \bibfnamefont [1]{#1}%
\providecommand \citenamefont [1]{#1}%
\providecommand \href@noop [0]{\@secondoftwo}%
\providecommand \href [0]{\begingroup \@sanitize@url \@href}%
\providecommand \@href[1]{\@@startlink{#1}\@@href}%
\providecommand \@@href[1]{\endgroup#1\@@endlink}%
\providecommand \@sanitize@url [0]{\catcode `\\12\catcode `\$12\catcode
  `\&12\catcode `\#12\catcode `\^12\catcode `\_12\catcode `\%12\relax}%
\providecommand \@@startlink[1]{}%
\providecommand \@@endlink[0]{}%
\providecommand \url  [0]{\begingroup\@sanitize@url \@url }%
\providecommand \@url [1]{\endgroup\@href {#1}{\urlprefix }}%
\providecommand \urlprefix  [0]{URL }%
\providecommand \Eprint [0]{\href }%
\providecommand \doibase [0]{http://dx.doi.org/}%
\providecommand \selectlanguage [0]{\@gobble}%
\providecommand \bibinfo  [0]{\@secondoftwo}%
\providecommand \bibfield  [0]{\@secondoftwo}%
\providecommand \translation [1]{[#1]}%
\providecommand \BibitemOpen [0]{}%
\providecommand \bibitemStop [0]{}%
\providecommand \bibitemNoStop [0]{.\EOS\space}%
\providecommand \EOS [0]{\spacefactor3000\relax}%
\providecommand \BibitemShut  [1]{\csname bibitem#1\endcsname}%
\let\auto@bib@innerbib\@empty
\bibitem [{\citenamefont {Troili}(1766)}]{Troili1766}%
  \BibitemOpen
  \bibfield  {author} {\bibinfo {author} {\bibfnamefont {D.}~\bibnamefont
  {Troili}},\ }\href@noop {} {\emph {\bibinfo {title} {Ragionamento della
  caduta di un sasso}}}\ (\bibinfo {year} {1766})\BibitemShut {NoStop}%
\bibitem [{\citenamefont {T\"{o}pel-Schadt}\ and\ \citenamefont
  {M\"{u}ller}(1982)}]{Muller1982}%
  \BibitemOpen
  \bibfield  {author} {\bibinfo {author} {\bibfnamefont {J.}~\bibnamefont
  {T\"{o}pel-Schadt}}\ and\ \bibinfo {author} {\bibfnamefont {W.~F.}\
  \bibnamefont {M\"{u}ller}},\ }\href {\doibase 10.1007/BF00308240} {\bibfield
  {journal} {\bibinfo  {journal} {Physics and Chemistry of Minerals}\ }\textbf
  {\bibinfo {volume} {8}},\ \bibinfo {pages} {175} (\bibinfo {year}
  {1982})}\BibitemShut {NoStop}%
\bibitem [{\citenamefont {Martin}\ \emph {et~al.}(2004)\citenamefont {Martin},
  \citenamefont {Vo\u{c}adlo}, \citenamefont {Alf\`e},\ and\ \citenamefont
  {Price}}]{Martin2004}%
  \BibitemOpen
  \bibfield  {author} {\bibinfo {author} {\bibfnamefont {P.}~\bibnamefont
  {Martin}}, \bibinfo {author} {\bibfnamefont {L.}~\bibnamefont {Vo\u{c}adlo}},
  \bibinfo {author} {\bibfnamefont {D.}~\bibnamefont {Alf\`e}}, \ and\ \bibinfo
  {author} {\bibfnamefont {G.~D.}\ \bibnamefont {Price}},\ }\href {\doibase 10}
  {\ \textbf {\bibinfo {volume} {68}},\ \bibinfo {pages} {813} (\bibinfo {year}
  {2004})}\BibitemShut {NoStop}%
\bibitem [{\citenamefont {Birch}(1952)}]{Birch1952}%
  \BibitemOpen
  \bibfield  {author} {\bibinfo {author} {\bibfnamefont {F.}~\bibnamefont
  {Birch}},\ }\href {\doibase 10.1029/JZ057i002p00227} {\bibfield  {journal}
  {\bibinfo  {journal} {Journal of Geophysical Research}\ }\textbf {\bibinfo
  {volume} {57}},\ \bibinfo {pages} {227} (\bibinfo {year} {1952})}\BibitemShut
  {NoStop}%
\bibitem [{\citenamefont {All\`egre}\ \emph {et~al.}(2001)\citenamefont
  {All\`egre}, \citenamefont {Manh\`es},\ and\ \citenamefont
  {Lewin}}]{Allegre2001}%
  \BibitemOpen
  \bibfield  {author} {\bibinfo {author} {\bibfnamefont {C.}~\bibnamefont
  {All\`egre}}, \bibinfo {author} {\bibfnamefont {G.}~\bibnamefont {Manh\`es}},
  \ and\ \bibinfo {author} {\bibfnamefont {E.}~\bibnamefont {Lewin}},\ }\href
  {\doibase 10.1016/S0012-821X(00)00359-9} {\bibfield  {journal} {\bibinfo
  {journal} {Earth and Planetary Science Letters}\ }\textbf {\bibinfo {volume}
  {185}},\ \bibinfo {pages} {49} (\bibinfo {year} {2001})}\BibitemShut
  {NoStop}%
\bibitem [{\citenamefont {Kusaba}\ \emph {et~al.}(1998)\citenamefont {Kusaba},
  \citenamefont {Syono}, \citenamefont {Kikegawa},\ and\ \citenamefont
  {Shimomura}}]{Kusaba1998}%
  \BibitemOpen
  \bibfield  {author} {\bibinfo {author} {\bibfnamefont {K.}~\bibnamefont
  {Kusaba}}, \bibinfo {author} {\bibfnamefont {Y.}~\bibnamefont {Syono}},
  \bibinfo {author} {\bibfnamefont {T.}~\bibnamefont {Kikegawa}}, \ and\
  \bibinfo {author} {\bibfnamefont {O.}~\bibnamefont {Shimomura}},\ }\href
  {\doibase http://dx.doi.org/10.1016/S0022-3697(98)00015-8} {\bibfield
  {journal} {\bibinfo  {journal} {Journal of Physics and Chemistry of Solids}\
  }\textbf {\bibinfo {volume} {59}},\ \bibinfo {pages} {945 } (\bibinfo {year}
  {1998})}\BibitemShut {NoStop}%
\bibitem [{\citenamefont {KUSABA}\ \emph {et~al.}(1997)\citenamefont {KUSABA},
  \citenamefont {SYONO}, \citenamefont {KIKEGAWA},\ and\ \citenamefont
  {SHIMOMURA}}]{Kusaba1997}%
  \BibitemOpen
  \bibfield  {author} {\bibinfo {author} {\bibfnamefont {K.}~\bibnamefont
  {KUSABA}}, \bibinfo {author} {\bibfnamefont {Y.}~\bibnamefont {SYONO}},
  \bibinfo {author} {\bibfnamefont {T.}~\bibnamefont {KIKEGAWA}}, \ and\
  \bibinfo {author} {\bibfnamefont {O.}~\bibnamefont {SHIMOMURA}},\ }\href
  {\doibase http://dx.doi.org/10.1016/S0022-3697(96)00120-5} {\bibfield
  {journal} {\bibinfo  {journal} {Journal of Physics and Chemistry of Solids}\
  }\textbf {\bibinfo {volume} {58}},\ \bibinfo {pages} {241 } (\bibinfo {year}
  {1997})}\BibitemShut {NoStop}%
\bibitem [{\citenamefont {Ohfuji}\ \emph {et~al.}(2007)\citenamefont {Ohfuji},
  \citenamefont {Sata}, \citenamefont {Kobayashi}, \citenamefont {Ohishi},
  \citenamefont {Hirose},\ and\ \citenamefont {Irifune}}]{ohfuji2007}%
  \BibitemOpen
  \bibfield  {author} {\bibinfo {author} {\bibfnamefont {H.}~\bibnamefont
  {Ohfuji}}, \bibinfo {author} {\bibfnamefont {N.}~\bibnamefont {Sata}},
  \bibinfo {author} {\bibfnamefont {H.}~\bibnamefont {Kobayashi}}, \bibinfo
  {author} {\bibfnamefont {Y.}~\bibnamefont {Ohishi}}, \bibinfo {author}
  {\bibfnamefont {K.}~\bibnamefont {Hirose}}, \ and\ \bibinfo {author}
  {\bibfnamefont {T.}~\bibnamefont {Irifune}},\ }\href {\doibase
  10.1007/s00269-007-0151-0} {\bibfield  {journal} {\bibinfo  {journal}
  {Physics and Chemistry of Minerals}\ }\textbf {\bibinfo {volume} {34}},\
  \bibinfo {pages} {335} (\bibinfo {year} {2007})}\BibitemShut {NoStop}%
\bibitem [{\citenamefont {Ono}\ \emph {et~al.}(2008)\citenamefont {Ono},
  \citenamefont {Oganov}, \citenamefont {Brodholt}, \citenamefont {Vo?adlo},
  \citenamefont {Wood}, \citenamefont {Lyakhov}, \citenamefont {Glass},
  \citenamefont {Côté},\ and\ \citenamefont {Price}}]{ono2008}%
  \BibitemOpen
  \bibfield  {author} {\bibinfo {author} {\bibfnamefont {S.}~\bibnamefont
  {Ono}}, \bibinfo {author} {\bibfnamefont {A.~R.}\ \bibnamefont {Oganov}},
  \bibinfo {author} {\bibfnamefont {J.~P.}\ \bibnamefont {Brodholt}}, \bibinfo
  {author} {\bibfnamefont {L.}~\bibnamefont {Vo?adlo}}, \bibinfo {author}
  {\bibfnamefont {I.~G.}\ \bibnamefont {Wood}}, \bibinfo {author}
  {\bibfnamefont {A.}~\bibnamefont {Lyakhov}}, \bibinfo {author} {\bibfnamefont
  {C.~W.}\ \bibnamefont {Glass}}, \bibinfo {author} {\bibfnamefont {A.~S.}\
  \bibnamefont {Côté}}, \ and\ \bibinfo {author} {\bibfnamefont {G.~D.}\
  \bibnamefont {Price}},\ }\href {\doibase
  http://dx.doi.org/10.1016/j.epsl.2008.05.017} {\bibfield  {journal} {\bibinfo
   {journal} {Earth and Planetary Science Letters}\ }\textbf {\bibinfo {volume}
  {272}},\ \bibinfo {pages} {481 } (\bibinfo {year} {2008})}\BibitemShut
  {NoStop}%
\bibitem [{\citenamefont {Adachi}\ and\ \citenamefont
  {Sato}(1968)}]{Adachi1968}%
  \BibitemOpen
  \bibfield  {author} {\bibinfo {author} {\bibfnamefont {K.}~\bibnamefont
  {Adachi}}\ and\ \bibinfo {author} {\bibfnamefont {K.}~\bibnamefont {Sato}},\
  }\href {\doibase http://dx.doi.org/10.1063/1.1656293} {\bibfield  {journal}
  {\bibinfo  {journal} {Journal of Applied Physics}\ }\textbf {\bibinfo
  {volume} {39}},\ \bibinfo {pages} {1343} (\bibinfo {year}
  {1968})}\BibitemShut {NoStop}%
\bibitem [{\citenamefont {Andresen}(1960)}]{Andresen1960}%
  \BibitemOpen
  \bibfield  {author} {\bibinfo {author} {\bibfnamefont {A.~F.}\ \bibnamefont
  {Andresen}},\ }\href {\doibase 10.3891/acta.chem.scand.14-0919} {\bibfield
  {journal} {\bibinfo  {journal} {Acta Chemica Scandinavica}\ }\textbf
  {\bibinfo {volume} {14}},\ \bibinfo {pages} {919} (\bibinfo {year}
  {1960})}\BibitemShut {NoStop}%
\bibitem [{\citenamefont {{E. F. Bertaut}}(1980)}]{Bertaut1980}%
  \BibitemOpen
  \bibfield  {author} {\bibinfo {author} {\bibnamefont {{E. F. Bertaut}}},\
  }\href {\doibase http://dx.doi.org/10.1351/pac198052010073} {\bibfield
  {journal} {\bibinfo  {journal} {Pure Appl. Chem.}\ }\textbf {\bibinfo
  {volume} {52}},\ \bibinfo {pages} {73} (\bibinfo {year} {1980})}\BibitemShut
  {NoStop}%
\bibitem [{\citenamefont {Bertaut}(1956)}]{Bertaut1956}%
  \BibitemOpen
  \bibfield  {author} {\bibinfo {author} {\bibfnamefont {E.~F.}\ \bibnamefont
  {Bertaut}},\ }\href@noop {} {\bibfield  {journal} {\bibinfo  {journal} {Bull.
  Soc. Fran\c{c}. Min\'er. Crist.}\ }\textbf {\bibinfo {volume} {79}},\
  \bibinfo {pages} {276} (\bibinfo {year} {1956})}\BibitemShut {NoStop}%
\bibitem [{\citenamefont {{E. F. Bertaut}}(1997)}]{Bertaut1997}%
  \BibitemOpen
  \bibfield  {author} {\bibinfo {author} {\bibnamefont {{E. F. Bertaut}}},\
  }\href {\doibase 10.1051/jp4:1997101} {\bibfield  {journal} {\bibinfo
  {journal} {J. Phys. IV France}\ }\textbf {\bibinfo {volume} {7}},\ \bibinfo
  {pages} {C1} (\bibinfo {year} {1997})}\BibitemShut {NoStop}%
\bibitem [{\citenamefont {van~den Berg}\ \emph {et~al.}(1969)\citenamefont
  {van~den Berg}, \citenamefont {van Delden},\ and\ \citenamefont
  {Bouman}}]{VandenBerg1969}%
  \BibitemOpen
  \bibfield  {author} {\bibinfo {author} {\bibfnamefont {C.~B.}\ \bibnamefont
  {van~den Berg}}, \bibinfo {author} {\bibfnamefont {J.~E.}\ \bibnamefont {van
  Delden}}, \ and\ \bibinfo {author} {\bibfnamefont {J.}~\bibnamefont
  {Bouman}},\ }\href {\doibase 10.1002/pssb.19690360249} {\bibfield  {journal}
  {\bibinfo  {journal} {physica status solidi (b)}\ }\textbf {\bibinfo {volume}
  {36}},\ \bibinfo {pages} {K89} (\bibinfo {year} {1969})}\BibitemShut
  {NoStop}%
\bibitem [{\citenamefont {van~den Berg}(1970)}]{VandenBerg1970}%
  \BibitemOpen
  \bibfield  {author} {\bibinfo {author} {\bibfnamefont {C.~B.}\ \bibnamefont
  {van~den Berg}},\ }\href {\doibase 10.1002/pssb.19700400243} {\bibfield
  {journal} {\bibinfo  {journal} {physica status solidi}\ }\textbf {\bibinfo
  {volume} {40}},\ \bibinfo {pages} {K65} (\bibinfo {year} {1970})}\BibitemShut
  {NoStop}%
\bibitem [{\citenamefont {Li}\ and\ \citenamefont {Franzen}(1996)}]{Li1996}%
  \BibitemOpen
  \bibfield  {author} {\bibinfo {author} {\bibfnamefont {F.}~\bibnamefont
  {Li}}\ and\ \bibinfo {author} {\bibfnamefont {H.~F.}\ \bibnamefont
  {Franzen}},\ }\href {\doibase 10.1016/0925-8388(96)02207-4} {\bibfield
  {journal} {\bibinfo  {journal} {Journal of Alloys and Compounds}\ }\textbf
  {\bibinfo {volume} {238}},\ \bibinfo {pages} {73} (\bibinfo {year}
  {1996})}\BibitemShut {NoStop}%
\bibitem [{\citenamefont {Eerenstein}\ \emph {et~al.}(2006)\citenamefont
  {Eerenstein}, \citenamefont {Mathur},\ and\ \citenamefont
  {Scott}}]{Scott2006}%
  \BibitemOpen
  \bibfield  {author} {\bibinfo {author} {\bibfnamefont {W.}~\bibnamefont
  {Eerenstein}}, \bibinfo {author} {\bibfnamefont {N.~D.}\ \bibnamefont
  {Mathur}}, \ and\ \bibinfo {author} {\bibfnamefont {J.~F.}\ \bibnamefont
  {Scott}},\ }\href {\doibase 10.1038/nature05023} {\bibfield  {journal}
  {\bibinfo  {journal} {Nature}\ }\textbf {\bibinfo {volume} {442}},\ \bibinfo
  {pages} {759} (\bibinfo {year} {2006})}\BibitemShut {NoStop}%
\bibitem [{\citenamefont {Khomskii}(2009)}]{Khomskii2009}%
  \BibitemOpen
  \bibfield  {author} {\bibinfo {author} {\bibfnamefont {D.}~\bibnamefont
  {Khomskii}},\ }\href {\doibase 10.1103/Physics.2.20} {\bibfield  {journal}
  {\bibinfo  {journal} {Physics}\ }\textbf {\bibinfo {volume} {2}},\ \bibinfo
  {pages} {20} (\bibinfo {year} {2009})}\BibitemShut {NoStop}%
\bibitem [{\citenamefont {Guenon}\ \emph {et~al.}(2015)\citenamefont {Guenon},
  \citenamefont {Ramirez}, \citenamefont {Basaran}, \citenamefont {Wampler},
  \citenamefont {Thiemens},\ and\ \citenamefont {Schuller}}]{Guenon2015}%
  \BibitemOpen
  \bibfield  {author} {\bibinfo {author} {\bibfnamefont {S.}~\bibnamefont
  {Guenon}}, \bibinfo {author} {\bibfnamefont {J.~G.}\ \bibnamefont {Ramirez}},
  \bibinfo {author} {\bibfnamefont {A.~C.}\ \bibnamefont {Basaran}}, \bibinfo
  {author} {\bibfnamefont {J.}~\bibnamefont {Wampler}}, \bibinfo {author}
  {\bibfnamefont {M.}~\bibnamefont {Thiemens}}, \ and\ \bibinfo {author}
  {\bibfnamefont {I.~K.}\ \bibnamefont {Schuller}},\ }\href@noop {} {\bibfield
  {journal} {\bibinfo  {journal} {arXiv}\ ,\ \bibinfo {pages} {1509.04452}}
  (\bibinfo {year} {2015})}\BibitemShut {NoStop}%
\bibitem [{\citenamefont {Cano}(2014)}]{Cano2014}%
  \BibitemOpen
  \bibfield  {author} {\bibinfo {author} {\bibfnamefont {A.}~\bibnamefont
  {Cano}},\ }\href {\doibase 10.1103/PhysRevB.89.214107} {\bibfield  {journal}
  {\bibinfo  {journal} {Phys. Rev. B}\ }\textbf {\bibinfo {volume} {89}},\
  \bibinfo {pages} {214107} (\bibinfo {year} {2014})}\BibitemShut {NoStop}%
\bibitem [{\citenamefont {Spaldin}\ \emph {et~al.}(2013)\citenamefont
  {Spaldin}, \citenamefont {Fechner}, \citenamefont {Bousquet}, \citenamefont
  {Balatsky},\ and\ \citenamefont {Nordstr\"om}}]{spaldin2013}%
  \BibitemOpen
  \bibfield  {author} {\bibinfo {author} {\bibfnamefont {N.~A.}\ \bibnamefont
  {Spaldin}}, \bibinfo {author} {\bibfnamefont {M.}~\bibnamefont {Fechner}},
  \bibinfo {author} {\bibfnamefont {E.}~\bibnamefont {Bousquet}}, \bibinfo
  {author} {\bibfnamefont {A.}~\bibnamefont {Balatsky}}, \ and\ \bibinfo
  {author} {\bibfnamefont {L.}~\bibnamefont {Nordstr\"om}},\ }\href {\doibase
  10.1103/PhysRevB.88.094429} {\bibfield  {journal} {\bibinfo  {journal} {Phys.
  Rev. B}\ }\textbf {\bibinfo {volume} {88}},\ \bibinfo {pages} {094429}
  (\bibinfo {year} {2013})}\BibitemShut {NoStop}%
\bibitem [{\citenamefont {Gosselin}\ \emph {et~al.}(1976)\citenamefont
  {Gosselin}, \citenamefont {Townsend},\ and\ \citenamefont
  {Tremblay}}]{Gosselin1976}%
  \BibitemOpen
  \bibfield  {author} {\bibinfo {author} {\bibfnamefont {J.~R.}\ \bibnamefont
  {Gosselin}}, \bibinfo {author} {\bibfnamefont {M.~G.}\ \bibnamefont
  {Townsend}}, \ and\ \bibinfo {author} {\bibfnamefont {R.~J.}\ \bibnamefont
  {Tremblay}},\ }\href {\doibase 10.1016/0038-1098(76)90922-4} {\bibfield
  {journal} {\bibinfo  {journal} {Solid State Communications}\ }\textbf
  {\bibinfo {volume} {19}},\ \bibinfo {pages} {799} (\bibinfo {year}
  {1976})}\BibitemShut {NoStop}%
\bibitem [{\citenamefont {Horwood}\ \emph {et~al.}(1976)\citenamefont
  {Horwood}, \citenamefont {Townsend},\ and\ \citenamefont
  {Webster}}]{Horwood1976}%
  \BibitemOpen
  \bibfield  {author} {\bibinfo {author} {\bibfnamefont {J.}~\bibnamefont
  {Horwood}}, \bibinfo {author} {\bibfnamefont {M.}~\bibnamefont {Townsend}}, \
  and\ \bibinfo {author} {\bibfnamefont {A.}~\bibnamefont {Webster}},\ }\href
  {\doibase http://dx.doi.org/10.1016/0022-4596(76)90198-5} {\bibfield
  {journal} {\bibinfo  {journal} {Journal of Solid State Chemistry}\ }\textbf
  {\bibinfo {volume} {17}},\ \bibinfo {pages} {35 } (\bibinfo {year}
  {1976})}\BibitemShut {NoStop}%
\bibitem [{\citenamefont {Antonov}\ \emph {et~al.}(2009)\citenamefont
  {Antonov}, \citenamefont {Bekenov}, \citenamefont {Shpak}, \citenamefont
  {Germash}, \citenamefont {Yaresko},\ and\ \citenamefont
  {Jepsen}}]{Antonov2009}%
  \BibitemOpen
  \bibfield  {author} {\bibinfo {author} {\bibfnamefont {V.~N.}\ \bibnamefont
  {Antonov}}, \bibinfo {author} {\bibfnamefont {L.~V.}\ \bibnamefont
  {Bekenov}}, \bibinfo {author} {\bibfnamefont {A.~P.}\ \bibnamefont {Shpak}},
  \bibinfo {author} {\bibfnamefont {L.~P.}\ \bibnamefont {Germash}}, \bibinfo
  {author} {\bibfnamefont {A.~N.}\ \bibnamefont {Yaresko}}, \ and\ \bibinfo
  {author} {\bibfnamefont {O.}~\bibnamefont {Jepsen}},\ }\href {\doibase
  http://dx.doi.org/10.1063/1.3269719} {\bibfield  {journal} {\bibinfo
  {journal} {Journal of Applied Physics}\ }\textbf {\bibinfo {volume} {106}},\
  \bibinfo {eid} {123907} (\bibinfo {year} {2009})}\BibitemShut {NoStop}%
\bibitem [{\citenamefont {Rohrbach}\ \emph {et~al.}(2003)\citenamefont
  {Rohrbach}, \citenamefont {Hafner},\ and\ \citenamefont
  {Kresse}}]{Rohrbach2003}%
  \BibitemOpen
  \bibfield  {author} {\bibinfo {author} {\bibfnamefont {A.}~\bibnamefont
  {Rohrbach}}, \bibinfo {author} {\bibfnamefont {J.}~\bibnamefont {Hafner}}, \
  and\ \bibinfo {author} {\bibfnamefont {G.}~\bibnamefont {Kresse}},\ }\href
  {http://stacks.iop.org/0953-8984/15/i=6/a=325} {\bibfield  {journal}
  {\bibinfo  {journal} {Journal of Physics: Condensed Matter}\ }\textbf
  {\bibinfo {volume} {15}},\ \bibinfo {pages} {979} (\bibinfo {year}
  {2003})}\BibitemShut {NoStop}%
\bibitem [{\citenamefont {Hobbs}\ and\ \citenamefont
  {Hafner}(1999)}]{Hobbs1999}%
  \BibitemOpen
  \bibfield  {author} {\bibinfo {author} {\bibfnamefont {D.}~\bibnamefont
  {Hobbs}}\ and\ \bibinfo {author} {\bibfnamefont {J.}~\bibnamefont {Hafner}},\
  }\href {http://stacks.iop.org/0953-8984/11/i=42/a=303} {\bibfield  {journal}
  {\bibinfo  {journal} {Journal of Physics: Condensed Matter}\ }\textbf
  {\bibinfo {volume} {11}},\ \bibinfo {pages} {8197} (\bibinfo {year}
  {1999})}\BibitemShut {NoStop}%
\bibitem [{\citenamefont {Chen}\ \emph {et~al.}(2014)\citenamefont {Chen},
  \citenamefont {Dai}, \citenamefont {Feng}, \citenamefont {Xiang},\ and\
  \citenamefont {Zhang}}]{Chen2014}%
  \BibitemOpen
  \bibfield  {author} {\bibinfo {author} {\bibfnamefont {X.}~\bibnamefont
  {Chen}}, \bibinfo {author} {\bibfnamefont {P.}~\bibnamefont {Dai}}, \bibinfo
  {author} {\bibfnamefont {D.}~\bibnamefont {Feng}}, \bibinfo {author}
  {\bibfnamefont {T.}~\bibnamefont {Xiang}}, \ and\ \bibinfo {author}
  {\bibfnamefont {F.-C.}\ \bibnamefont {Zhang}},\ }\href {\doibase
  10.1093/nsr/nwu007} {\ \textbf {\bibinfo {volume} {1}},\ \bibinfo {pages}
  {371} (\bibinfo {year} {2014})}\BibitemShut {NoStop}%
\bibitem [{\citenamefont {Salje}\ \emph {et~al.}(2005)\citenamefont {Salje},
  \citenamefont {Alexandrov},\ and\ \citenamefont {Liang}}]{Salje}%
  \BibitemOpen
  \bibfield  {author} {\bibinfo {author} {\bibfnamefont {E.~K.~H.}\
  \bibnamefont {Salje}}, \bibinfo {author} {\bibfnamefont {A.~S.}\ \bibnamefont
  {Alexandrov}}, \ and\ \bibinfo {author} {\bibfnamefont {W.~Y.}\ \bibnamefont
  {Liang}},\ }\href@noop {} {\emph {\bibinfo {title} {Polarons and Bipolarons
  in High-Tc Superconductors and Related Materials}}}\ (\bibinfo  {publisher}
  {Cambridge University Press},\ \bibinfo {year} {2005})\BibitemShut {NoStop}%
\bibitem [{\citenamefont {Fiebig}(2005)}]{fiebig2005}%
  \BibitemOpen
  \bibfield  {author} {\bibinfo {author} {\bibfnamefont {M.}~\bibnamefont
  {Fiebig}},\ }\href {\doibase 10.1088/0022-3727/38/8/R01} {\bibfield
  {journal} {\bibinfo  {journal} {J. Phys. D: Appl. Phys.}\ }\textbf {\bibinfo
  {volume} {38}},\ \bibinfo {pages} {R123} (\bibinfo {year}
  {2005})}\BibitemShut {NoStop}%
\bibitem [{\citenamefont {Bousquet}\ \emph {et~al.}(2011)\citenamefont
  {Bousquet}, \citenamefont {Spaldin},\ and\ \citenamefont
  {Delaney}}]{bousquet2011}%
  \BibitemOpen
  \bibfield  {author} {\bibinfo {author} {\bibfnamefont {E.}~\bibnamefont
  {Bousquet}}, \bibinfo {author} {\bibfnamefont {N.~A.}\ \bibnamefont
  {Spaldin}}, \ and\ \bibinfo {author} {\bibfnamefont {K.~T.}\ \bibnamefont
  {Delaney}},\ }\href {\doibase 10.1103/PhysRevLett.106.107202} {\bibfield
  {journal} {\bibinfo  {journal} {Phys. Rev. Lett.}\ }\textbf {\bibinfo
  {volume} {106}},\ \bibinfo {pages} {107202} (\bibinfo {year}
  {2011})}\BibitemShut {NoStop}%
\bibitem [{\citenamefont {\'I\~niguez}(2008)}]{iniguez2008}%
  \BibitemOpen
  \bibfield  {author} {\bibinfo {author} {\bibfnamefont {J.}~\bibnamefont
  {\'I\~niguez}},\ }\href {\doibase 10.1103/PhysRevLett.101.117201} {\bibfield
  {journal} {\bibinfo  {journal} {Phys. Rev. Lett.}\ }\textbf {\bibinfo
  {volume} {101}},\ \bibinfo {pages} {117201} (\bibinfo {year}
  {2008})}\BibitemShut {NoStop}%
\bibitem [{\citenamefont {Malashevich}\ \emph {et~al.}(2012)\citenamefont
  {Malashevich}, \citenamefont {Coh}, \citenamefont {Souza},\ and\
  \citenamefont {Vanderbilt}}]{malashevich2012}%
  \BibitemOpen
  \bibfield  {author} {\bibinfo {author} {\bibfnamefont {A.}~\bibnamefont
  {Malashevich}}, \bibinfo {author} {\bibfnamefont {S.}~\bibnamefont {Coh}},
  \bibinfo {author} {\bibfnamefont {I.}~\bibnamefont {Souza}}, \ and\ \bibinfo
  {author} {\bibfnamefont {D.}~\bibnamefont {Vanderbilt}},\ }\href {\doibase
  10.1103/PhysRevB.86.094430} {\bibfield  {journal} {\bibinfo  {journal} {Phys.
  Rev. B}\ }\textbf {\bibinfo {volume} {86}},\ \bibinfo {pages} {094430}
  (\bibinfo {year} {2012})}\BibitemShut {NoStop}%
\bibitem [{\citenamefont {Ye}\ and\ \citenamefont
  {Vanderbilt}(2014)}]{Vanderbilt2014}%
  \BibitemOpen
  \bibfield  {author} {\bibinfo {author} {\bibfnamefont {M.}~\bibnamefont
  {Ye}}\ and\ \bibinfo {author} {\bibfnamefont {D.}~\bibnamefont
  {Vanderbilt}},\ }\href {\doibase 10.1103/PhysRevB.89.064301} {\bibfield
  {journal} {\bibinfo  {journal} {Phys. Rev. B}\ }\textbf {\bibinfo {volume}
  {89}},\ \bibinfo {pages} {064301} (\bibinfo {year} {2014})}\BibitemShut
  {NoStop}%
\bibitem [{\citenamefont {Orobengoa}\ \emph {et~al.}(2009)\citenamefont
  {Orobengoa}, \citenamefont {Capillas}, \citenamefont {Aroyo},\ and\
  \citenamefont {Perez-Mato}}]{amplimode}%
  \BibitemOpen
  \bibfield  {author} {\bibinfo {author} {\bibfnamefont {D.}~\bibnamefont
  {Orobengoa}}, \bibinfo {author} {\bibfnamefont {C.}~\bibnamefont {Capillas}},
  \bibinfo {author} {\bibfnamefont {M.~I.}\ \bibnamefont {Aroyo}}, \ and\
  \bibinfo {author} {\bibfnamefont {J.~M.}\ \bibnamefont {Perez-Mato}},\ }\href
  {\doibase 10.1107/S0021889809028064} {\bibfield  {journal} {\bibinfo
  {journal} {Journal of Applied Crystallography}\ }\textbf {\bibinfo {volume}
  {42}},\ \bibinfo {pages} {820} (\bibinfo {year} {2009})}\BibitemShut
  {NoStop}%
\bibitem [{\citenamefont {Fennie}\ and\ \citenamefont
  {Rabe}(2005)}]{Fennie2005}%
  \BibitemOpen
  \bibfield  {author} {\bibinfo {author} {\bibfnamefont {C.~J.}\ \bibnamefont
  {Fennie}}\ and\ \bibinfo {author} {\bibfnamefont {K.~M.}\ \bibnamefont
  {Rabe}},\ }\href {\doibase 10.1103/PhysRevB.72.100103} {\bibfield  {journal}
  {\bibinfo  {journal} {Phys. Rev. B}\ }\textbf {\bibinfo {volume} {72}},\
  \bibinfo {pages} {100103} (\bibinfo {year} {2005})}\BibitemShut {NoStop}%
\bibitem [{\citenamefont {Shirodkar}\ and\ \citenamefont
  {Waghmare}(2014)}]{Shirodkar2014}%
  \BibitemOpen
  \bibfield  {author} {\bibinfo {author} {\bibfnamefont {S.~N.}\ \bibnamefont
  {Shirodkar}}\ and\ \bibinfo {author} {\bibfnamefont {U.~V.}\ \bibnamefont
  {Waghmare}},\ }\href {\doibase 10.1103/PhysRevLett.112.157601} {\bibfield
  {journal} {\bibinfo  {journal} {Phys. Rev. Lett.}\ }\textbf {\bibinfo
  {volume} {112}},\ \bibinfo {pages} {157601} (\bibinfo {year}
  {2014})}\BibitemShut {NoStop}%
\bibitem [{\citenamefont {Fiebig}\ and\ \citenamefont
  {Spaldin}(2009)}]{Fiebig2009}%
  \BibitemOpen
  \bibfield  {author} {\bibinfo {author} {\bibfnamefont {M.}~\bibnamefont
  {Fiebig}}\ and\ \bibinfo {author} {\bibfnamefont {N.~A.}\ \bibnamefont
  {Spaldin}},\ }\href {\doibase 10.1140/epjb/e2009-00266-4} {\bibfield
  {journal} {\bibinfo  {journal} {The European Physical Journal B}\ }\textbf
  {\bibinfo {volume} {71}},\ \bibinfo {pages} {293} (\bibinfo {year}
  {2009})}\BibitemShut {NoStop}%
\bibitem [{\citenamefont {Gonze}\ and\ \citenamefont {{\em et
  al.}}(2009)}]{Gonze2009}%
  \BibitemOpen
  \bibfield  {author} {\bibinfo {author} {\bibfnamefont {X.}~\bibnamefont
  {Gonze}}\ and\ \bibinfo {author} {\bibnamefont {{\em et al.}}},\ }\href
  {\doibase 10.1016/j.cpc.2009.07.007} {\bibfield  {journal} {\bibinfo
  {journal} {Computer Physics Communications}\ }\textbf {\bibinfo {volume}
  {180}},\ \bibinfo {pages} {2582} (\bibinfo {year} {2009})}\BibitemShut
  {NoStop}%
\bibitem [{\citenamefont {Kresse}\ and\ \citenamefont
  {Furthm{\"u}ller}(1996)}]{vasp1}%
  \BibitemOpen
  \bibfield  {author} {\bibinfo {author} {\bibfnamefont {G.}~\bibnamefont
  {Kresse}}\ and\ \bibinfo {author} {\bibfnamefont {J.}~\bibnamefont
  {Furthm{\"u}ller}},\ }\href {\doibase 10.1103/PhysRevB.54.11169} {\bibfield
  {journal} {\bibinfo  {journal} {Phys. Rev. B}\ }\textbf {\bibinfo {volume}
  {54}},\ \bibinfo {pages} {11169} (\bibinfo {year} {1996})}\BibitemShut
  {NoStop}%
\bibitem [{\citenamefont {Kresse}\ and\ \citenamefont
  {Joubert}(1999)}]{kresse1999}%
  \BibitemOpen
  \bibfield  {author} {\bibinfo {author} {\bibfnamefont {G.}~\bibnamefont
  {Kresse}}\ and\ \bibinfo {author} {\bibfnamefont {D.}~\bibnamefont
  {Joubert}},\ }\href {\doibase 10.1103/PhysRevB.59.1758} {\bibfield  {journal}
  {\bibinfo  {journal} {Phys. Rev. B}\ }\textbf {\bibinfo {volume} {59}},\
  \bibinfo {pages} {1758} (\bibinfo {year} {1999})}\BibitemShut {NoStop}%
\bibitem [{\citenamefont {Perdew}\ \emph {et~al.}(1996)\citenamefont {Perdew},
  \citenamefont {Burke},\ and\ \citenamefont {Ernzerhof}}]{Perdew1996}%
  \BibitemOpen
  \bibfield  {author} {\bibinfo {author} {\bibfnamefont {J.~P.}\ \bibnamefont
  {Perdew}}, \bibinfo {author} {\bibfnamefont {K.}~\bibnamefont {Burke}}, \
  and\ \bibinfo {author} {\bibfnamefont {M.}~\bibnamefont {Ernzerhof}},\ }\href
  {\doibase 10.1103/PhysRevLett.77.3865} {\bibfield  {journal} {\bibinfo
  {journal} {Phys. Rev. Lett.}\ }\textbf {\bibinfo {volume} {77}},\ \bibinfo
  {pages} {3865} (\bibinfo {year} {1996})}\BibitemShut {NoStop}%
\bibitem [{\citenamefont {Perdew}\ \emph {et~al.}(2008)\citenamefont {Perdew},
  \citenamefont {Ruzsinszky}, \citenamefont {Csonka}, \citenamefont {Vydrov},
  \citenamefont {Scuseria}, \citenamefont {Constantin}, \citenamefont {Zhou},\
  and\ \citenamefont {Burke}}]{Perdew2008}%
  \BibitemOpen
  \bibfield  {author} {\bibinfo {author} {\bibfnamefont {J.~P.}\ \bibnamefont
  {Perdew}}, \bibinfo {author} {\bibfnamefont {A.}~\bibnamefont {Ruzsinszky}},
  \bibinfo {author} {\bibfnamefont {G.~I.}\ \bibnamefont {Csonka}}, \bibinfo
  {author} {\bibfnamefont {O.~A.}\ \bibnamefont {Vydrov}}, \bibinfo {author}
  {\bibfnamefont {G.~E.}\ \bibnamefont {Scuseria}}, \bibinfo {author}
  {\bibfnamefont {L.~A.}\ \bibnamefont {Constantin}}, \bibinfo {author}
  {\bibfnamefont {X.}~\bibnamefont {Zhou}}, \ and\ \bibinfo {author}
  {\bibfnamefont {K.}~\bibnamefont {Burke}},\ }\href {\doibase
  10.1103/PhysRevLett.100.136406} {\bibfield  {journal} {\bibinfo  {journal}
  {Phys. Rev. Lett.}\ }\textbf {\bibinfo {volume} {100}},\ \bibinfo {pages}
  {136406} (\bibinfo {year} {2008})}\BibitemShut {NoStop}%
\bibitem [{\citenamefont {Liechtenstein}\ \emph {et~al.}(1995)\citenamefont
  {Liechtenstein}, \citenamefont {Anisimov},\ and\ \citenamefont
  {Zaanen}}]{Liechtenstein1995}%
  \BibitemOpen
  \bibfield  {author} {\bibinfo {author} {\bibfnamefont {A.~I.}\ \bibnamefont
  {Liechtenstein}}, \bibinfo {author} {\bibfnamefont {V.~I.}\ \bibnamefont
  {Anisimov}}, \ and\ \bibinfo {author} {\bibfnamefont {J.}~\bibnamefont
  {Zaanen}},\ }\href {\doibase 10.1103/PhysRevB.52.R5467} {\bibfield  {journal}
  {\bibinfo  {journal} {Phys. Rev. B}\ }\textbf {\bibinfo {volume} {52}},\
  \bibinfo {pages} {R5467} (\bibinfo {year} {1995})}\BibitemShut {NoStop}%
\bibitem [{\citenamefont {Torrent}\ \emph {et~al.}(2008)\citenamefont
  {Torrent}, \citenamefont {Jollet}, \citenamefont {Bottin}, \citenamefont
  {Z\'erah},\ and\ \citenamefont {Gonze}}]{Torrent2008}%
  \BibitemOpen
  \bibfield  {author} {\bibinfo {author} {\bibfnamefont {M.}~\bibnamefont
  {Torrent}}, \bibinfo {author} {\bibfnamefont {F.}~\bibnamefont {Jollet}},
  \bibinfo {author} {\bibfnamefont {F.}~\bibnamefont {Bottin}}, \bibinfo
  {author} {\bibfnamefont {G.}~\bibnamefont {Z\'erah}}, \ and\ \bibinfo
  {author} {\bibfnamefont {X.}~\bibnamefont {Gonze}},\ }\href {\doibase
  10.1016/j.commatsci.2007.07.020} {\bibfield  {journal} {\bibinfo  {journal}
  {Computational Materials Science}\ }\textbf {\bibinfo {volume} {42}},\
  \bibinfo {pages} {337} (\bibinfo {year} {2008})}\BibitemShut {NoStop}%
\bibitem [{\citenamefont {Monkhorst}\ and\ \citenamefont
  {Pack}(1976)}]{Monkhorst1976}%
  \BibitemOpen
  \bibfield  {author} {\bibinfo {author} {\bibfnamefont {H.~J.}\ \bibnamefont
  {Monkhorst}}\ and\ \bibinfo {author} {\bibfnamefont {J.~D.}\ \bibnamefont
  {Pack}},\ }\href {\doibase 10.1103/PhysRevB.13.5188} {\bibfield  {journal}
  {\bibinfo  {journal} {Phys. Rev. B}\ }\textbf {\bibinfo {volume} {13}},\
  \bibinfo {pages} {5188} (\bibinfo {year} {1976})}\BibitemShut {NoStop}%
\bibitem [{\citenamefont {Gonze}(1995)}]{gonze1995b}%
  \BibitemOpen
  \bibfield  {author} {\bibinfo {author} {\bibfnamefont {X.}~\bibnamefont
  {Gonze}},\ }\href {\doibase 10.1103/PhysRevA.52.1096} {\bibfield  {journal}
  {\bibinfo  {journal} {Phys. Rev. A}\ }\textbf {\bibinfo {volume} {52}},\
  \bibinfo {pages} {1096} (\bibinfo {year} {1995})}\BibitemShut {NoStop}%
\bibitem [{\citenamefont {Gonze}\ and\ \citenamefont {Lee}(1997)}]{gonze1997}%
  \BibitemOpen
  \bibfield  {author} {\bibinfo {author} {\bibfnamefont {X.}~\bibnamefont
  {Gonze}}\ and\ \bibinfo {author} {\bibfnamefont {C.}~\bibnamefont {Lee}},\
  }\href {\doibase 10.1103/PhysRevB.55.10355} {\bibfield  {journal} {\bibinfo
  {journal} {Phys. Rev. B}\ }\textbf {\bibinfo {volume} {55}},\ \bibinfo
  {pages} {10355} (\bibinfo {year} {1997})}\BibitemShut {NoStop}%
\bibitem [{\citenamefont {King-Smith}\ and\ \citenamefont
  {Vanderbilt}(1993)}]{kingsmith1993}%
  \BibitemOpen
  \bibfield  {author} {\bibinfo {author} {\bibfnamefont {R.~D.}\ \bibnamefont
  {King-Smith}}\ and\ \bibinfo {author} {\bibfnamefont {D.}~\bibnamefont
  {Vanderbilt}},\ }\href {\doibase 10.1103/PhysRevB.47.1651} {\bibfield
  {journal} {\bibinfo  {journal} {Phys. Rev. B}\ }\textbf {\bibinfo {volume}
  {47}},\ \bibinfo {pages} {1651} (\bibinfo {year} {1993})}\BibitemShut
  {NoStop}%
\bibitem [{\citenamefont {Bousquet}\ and\ \citenamefont
  {Spaldin}(2011)}]{bousquet2011b}%
  \BibitemOpen
  \bibfield  {author} {\bibinfo {author} {\bibfnamefont {E.}~\bibnamefont
  {Bousquet}}\ and\ \bibinfo {author} {\bibfnamefont {N.~A.}\ \bibnamefont
  {Spaldin}},\ }\href {\doibase 10.1103/PhysRevLett.107.197603} {\bibfield
  {journal} {\bibinfo  {journal} {Phys. Rev. Lett.}\ }\textbf {\bibinfo
  {volume} {107}},\ \bibinfo {pages} {197603} (\bibinfo {year}
  {2011})}\BibitemShut {NoStop}%
\end{thebibliography}
\end{document}